\documentstyle[prl,aps,epsf,multicol]{revtex}

\begin{document}
\draft
\tighten

\title{Viscoelasticity from a Microscopic Model of Dislocation Dynamics}

\author{M. Cristina Marchetti, Karl Saunders}
\address{Physics Department,Syracuse University, Syracuse, NY 13244}

\date{\today}
\maketitle
\begin{abstract}
  
It is shown  that the dynamics of a 
two-dimensional crystal with a finite concentration of dislocations, as well 
as vacancy and interstitial defects, is governed by the hydrodynamic equations 
of a viscoelastic medium. At the longest length scales the viscoelasticity is 
described by the simplest {\em Maxwell} model, whose shear and compressional 
relaxation times are obtained in terms of microscopic quantities, including 
the density of free dislocations. At short length scales,
bond orientational order effects become 
important and lead to wavevector dependent corrections to  
the relaxation times.  

\end{abstract}
\pacs{83.60.Bc,62.20.Fe}

\vspace{-0.5cm}
\begin{multicols}{2}
\narrowtext

\section{Introduction}
Experimental \cite{Bhatt,matsuda96,abulafia96,Hend} and numerical \cite{SB,FMM,GBD96,nori96,olson97} studies of the 
dynamics of driven disordered media, such as vortex arrays in type-II 
superconductors and charge density waves in metals, have clearly indicated 
that when the disorder is strong these systems
exhibit a spatially inhomogeneous plastic response  upon depinning,
without long wavelength elastic restoring forces \cite{FisherReview}. 
In this plastic regime
dislocations proliferate due to the interplay of drive, disorder and 
interactions, and the system is broken up in fluid-like regions flowing around
solid regions \cite{FMM}. It was proposed recently \cite{MMP} that a  description
of shear deformations in this plastic regime may be obtained by 
focusing on the dynamics of coarse-grained degrees of freedom 
(the solid-like regions) and
replacing the
elastic couplings of local displacements with viscoelastic 
couplings of local velocities. It was found that the simplest, Maxwell, model
of viscoelasticity yields interesting features, such hysteretic depinning and
switching, which have also been observed in experiments. The local
viscoelastic couplings were introduced as an ad hoc way of mimicking
the presence in the system at any given time of unbound dislocations,
whose motion is in turn responsible for plastic slip. 

In this paper we provide some justification to the model
studied in Ref. \onlinecite{MMP} by showing that the dynamics of 
a two-dimensional crystal with a finite
concentration of {\em annealed} free dislocations, as well as vacancy and
interstitial defects, is governed by the hydrodynamic equations
of a viscoelastic medium (with hexatic order). Starting with the hydrodynamic equations for
a two-dimensional solid with finite concentrations of dislocations, vacancies
and interstitials obtained some time ago by Zippelius, Halperin and 
Nelson \cite{ZHN} (ZHN), we show that such equations can be recast in the form 
of hydrodynamic equations for a viscoelastic medium, with a Maxwell form 
for the nonlocal transport coefficients \cite{Frenkel}. 

It has of course long been recognized
that the creation and motion of dislocations are the main mechanism by which 
a crystal undergoes plastic deformations 
\cite{taylor,Cottrell}. A number of efforts \cite{BHZ,DSP} have been made to 
describe the motion of dislocations through crystalline solids
and relate the plastic strain rate to the dislocation dynamics.
There has also been work specifically on the effect of dislocation
motion on linear and nonlinear stress relaxation in a two-dimensional crystal
\cite{BHZ}, where it was shown {\em for a particular geometry} that
the crystal responds viscoelastically when free dislocations are present.
Our work generalizes this by obtaining equations that describe the 
viscoelastic response of a two-dimensional solid to stresses 
in an {\em arbitrary geometry}. The equations incorporate viscoelastic effects 
in the response to both shear and compressional 
deformations, and have precisely the structure of the 
viscoelastic equations for a simple viscous fluid \cite{BoonYip}.
Furthermore, our derivation yields expressions for the (Maxwell) 
relaxation times for shear and compression in terms of microscopic 
parameters.  

Of course, in the presence of {\em quenched} disorder
and drive, dislocations are continuously generated and healed in the system.
Furthermore, they are not always free to move and relax a local stress as
they may be pinned by disorder. 
For these reasons one may question the assumption 
behind the model introduced in Ref. \onlinecite{MMP}, namely that the plastic response 
of extended media driven over strong quenched disorder 
at zero temperature may be described at large scale
by the same equation that govern the stress relaxation 
due to the motion of annealed
dislocations. At best, the effective density of ``free''
dislocations that can relax local stresses will be a strong function 
of the applied driving force, disorder, and possibly time. 
As shown in Ref. \onlinecite{MMP}, the depinning of a driven viscoelastic medium
does, however, exhibit a number of features seen in experiments,
namely memory effects and coexistence of weakly and strongly 
pinned degrees of freedom. It does, therefore provide a useful starting
model for the description of the complex dynamics of these systems.

Finally, it should be pointed out that crystals with annealed dislocations 
and simple viscous fluids are not the only examples of viscoelastic media. 
Linear viscoelasticity is of course one of the distinctive properties of 
complex and polymeric fluids, that can fill  containers of any shape 
and yet may shrink like rubber when stretched and released quickly. 
Molecular theories of viscoelasticity for complex fluids 
have been developed by various methods and there is a vast literature
on the subject \cite{DoiEdwards}.

\section{Summary of Main Results and Outline}
\label{summary}

It is well known that  a dense fluid has insufficient time
to flow in response to a high frequency strain rate, but instead 
reacts elastically, as a solid would \cite{HansenMcDonald}. This leads to the appearance 
in the fluid of propagating shear waves, with an associated peak at 
nonzero frequency in the spectrum of transverse current fluctuations. 
Such viscoelastic effects are easily incorporated phenomenologically into
the hydrodynamic description of fluids \cite{Frenkel}.
If a shearing force is applied to a fluid 
yielding a stress $\sigma_{xy}$, the local strain at each point can 
be expressed in terms of derivatives of the displacement field
${\bf u}$ at that point. 
In a steady-state situation the flow
is purely viscous and the stress is proportional to 
the local rate of strain,
\begin{equation}
\sigma_{xy}=\eta\frac{\partial}{\partial t}\Big(\frac{\partial u_x}{\partial y}
   +\frac{\partial u_y}{\partial x}\Big)\;,
\end{equation}
where $\eta$ is the shear viscosity. 
By contrast, if the force is applied suddenly,
the instantaneous displacement is related to the stress 
via a typical elastic stress-strain 
relation,
\begin{equation}
\sigma_{xy}=\mu\Big(\frac{\partial{u}_x}{\partial y}
   +\frac{\partial{u}_y}{\partial x}\Big)\;,
\end{equation}
where $\mu$ is the instantaneous (high-frequency) shear modulus.
An interpolation between these two forms yields the Maxwell model
of viscoelasticity, where the stress-strain rate relation takes the form of
a differential equation,
\begin{equation}
\label{Maxwell}
\frac{\partial\sigma_{xy}}{\partial t}+\frac{\sigma_{xy}}{\tau_M}=
   \mu\Big(\frac{\partial v_x}{\partial y}
   +\frac{\partial v_y}{\partial x}\Big)\;.
\end{equation}
The local flow velocity, ${\bf v}$, of the constituent particles is identified 
as the rate of change of the displacement, ${\bf v}=\partial_t{\bf u}$,
and $\tau_M=\eta/\mu$ is called the Maxwell relaxation time. 
The Maxwell model of viscoelasticity given by Eq. (\ref{Maxwell}) yields 
elastic response
to high frequency shearing forces ($\omega\tau_M>>1$)
and viscous response at low frequency ($\omega\tau_M<<1$).

In general, viscoelastic effects are present in the response 
to both shear and compression and can be described by modifying the 
hydrodynamic equations of a fluid by introducing nonlocal
(in space and time) thermodynamic quantities and transport coefficients.
This can be done phenomenologically through the use of frequency sum rules
as constraints, or can be formulated more systematically 
by deriving such equations via  formal projection operator methods 
or kinetic theory techniques \cite{BoonYip,HansenMcDonald}.
In its simplest realization this approach yields the Maxwell model
of viscoelasticity. 

In this paper we show that 
the hydrodynamic equations for a two-dimensional solid
with a finite concentration of free dislocations proposed by ZHN
are equivalent to a set of hydrodynamic equations for a simple 
fluid (with hexatic order), with a Maxwell-type stress-rate of 
strain relation. For convenience, we define a symmetrized strain 
rate, $v_{ij}$, as
\begin{eqnarray}
v_{ij}& \equiv &(\partial_i v_j + \partial_j v_i)/2\nonumber\\
&=& \tilde{v}_{ij}+\frac{1}{2}\delta_{ij}v_{kk}\;,
\label{v_ij}
\end{eqnarray}
where we have separated the traceless part $\tilde{v}_{ij}$ 
and the trace $v_{kk}=\nabla\cdot{\bf v}$.
It is also useful to separate the stress tensor
$\sigma_{ij}$ into its traceless part and its trace, i.e., 
\begin{equation}
\label{stress}
\sigma_{ij}=\tilde{\sigma}_{ij}+\frac{1}{2}\delta_{ij}\sigma_{kk}\;.
\end{equation}
We find that, at long wavelengths, the traceless part of 
the stress tensor is  related to the strain rate according to
\begin{eqnarray}
{\partial {\tilde\sigma}_{ij} \over \partial t} + {{\tilde\sigma}_{ij} \over \tau_s}=2\mu{\tilde v}_{ij}\;,
\label{sigmatilde_ij}
\end{eqnarray}
where $\tau_s$ is the shear relaxation time, which is obtained 
in terms of microscopic parameters (see Eq.~(\ref{tau_s}) below). 
Equation (\ref{sigmatilde_ij})
is precisely of the Maxwell form. If the product $\mu\tau_s$ 
is identified with the shear viscosity $\eta$ of the fluid, Eq.~(\ref{sigmatilde_ij})
yields viscous flow in response to low frequency shearing forces.
It also describes the fact that
a liquid always flows in response to a steady shear.

In contrast, even a liquid has a finite low
frequency compressional modulus, that is, it resists elastically low
frequency compressional deformations.
The trace of the stress tensor, $\sigma_{kk}$, contains both
reversible contributions proportional to pressure changes 
arising (assuming temperature
fluctuations are negligible) from the fractional change in volume
and irreversible and dissipative contributions proportional to 
velocity gradients. It can be written as
\begin{eqnarray}
\sigma_{kk} = -2 B_0 {\delta n\over n_0}+ \sigma_{kk}^{\rm dis}\;,
\label{sigma_kk}
\end{eqnarray}
with $\delta n$ the deviation of the density from its equilibrium value
$n_0$, and $B_0$  the low frequency
bulk modulus of the solid (which generally depends on the density
of vacancies and interstitials defects).
To a good approximation $B_0$ is also the compressibility of the liquid phase \cite{NH}. 
We find that the dissipative part $\sigma_{kk}^{\rm dis}$ of the stress tensor 
is related to the local rate of strain via the equation
\begin{eqnarray}
{\partial \sigma_{kk}^{\rm dis} \over \partial t} + {\sigma_{kk}^{\rm dis} \over \tau_b}=-2(B_\infty-B_0){\delta n \over n_0}\;,
\label{sigma_kk^ve}
\end{eqnarray}
where $B_\infty$ is the high frequency compressional modulus of the solid 
in the absence of defects. At high frequency defects do not have time to
respond to an applied force and therefore do not contribute to
the elastic constants. Finally,
$\tau_b$ is the compressional relaxation time, 
which again is given below
in terms of microscopic quantities. Both relaxation times, 
$\tau_s$ and $\tau_b$ 
are inversely proportional to the density of free dislocations, 
$n_f$. As a result, they both diverge as the melting temperature 
$T_m$ of the two-dimensional solid is approached from above as
\cite{NH,Y}
\begin{eqnarray}
\tau_{s,b} \propto \exp(b/(T-T_m)^\alpha)\;,
\label{tau_divergence}
\end{eqnarray}
with $\alpha=0.36963$.
At the longest length scales, the bulk viscoelasticity described by 
Eq.~(\ref{sigma_kk^ve}) 
is due solely to dislocation {\em climb}, 
while the shear viscoelasticity is due to both 
dislocation climb and {\em glide}. 
Dislocation climb which requires the creation of 
vacancies and interstitials, proceeds more slowly than dislocation glide which 
does not involve a change in particle number. 
For this reason, $\tau_b \gg \tau_s$, in 
general. 

Of course, the 2d solid does not melt directly into a isotropic liquid at
$T_m$, rather it
enters the hexatic phase, where it has only short-ranged translational order,
but retains quasi-long-ranged bond orientational order (BOO). At the 
longest length scales, the presence of BOO does not affect
the stress-rate of 
strain relations. When terms of higher order in the gradients of 
the hydrodynamic fields are included, the existence of BOO becomes 
apparent and modifies the stress-strain rate relations at finite length scales,
as discussed in Section \ref{grad2} below. In particular, this gives
wavevector-dependent corrections to $\tau_s$ and $\tau_b$,
corresponding to a distribution 
of relaxation times.

The rest of the paper is organized as follows. In Section \ref{review}
we review the hydrodynamic equations for a two-dimensional solid with 
vacancy and interstitial defects \cite{ZHN}.
In Section \ref{grad1} 
the viscoelastic stress-rate of strain relations
are derived in the longest length scale regime, and expressions for the 
long-wavelength
viscoelastic memory times are presented. In Section \ref{grad2}
the effects of incorporating higher order gradients are investigated, 
including the wavevector dependent corrections to $\tau_s$. 

\section{ZHN Hydrodynamics of a Solid with Dislocations} 
\label{review}

We begin by reviewing the hydrodynamics of the solid phase, 
following Refs.~\onlinecite{ZHN}, \onlinecite{martin72} and \onlinecite{landau}.
We consider a two-dimensional hexagonal crystal 
with a finite density of vacancies and interstitials.
As pointed out by Martin \cite{martin72}, incorporating a finite density of point defects
is crucial to properly distinguish between local lattice deformations
and local changes in the number density.
For simplicity, temperature fluctuations are neglected in all of the following.

In the absence of dislocations, the equations for conservation of 
density, $n({\bf r},t)$, 
momentum, ${\bf g}({\bf r},t)$, and net defect density, 
$n_{\Delta}({\bf r},t)$, i.e., density of interstitials minus 
density of vacancies, and the equation of motion for 
the displacement field, 
${\bf u}({\bf r},t)$, can be written as
\begin{mathletters}
\begin{eqnarray}
& &{\partial n \over \partial t} = -{1\over m} {\bf \nabla} \cdot {\bf g}\;,
\label{n}\\ 
& & {\partial g_i \over \partial t} = \partial_j \sigma_{ij}\;,
\label{g}\\ 
& &{\partial n_{\Delta} \over \partial t} = -{\bf \nabla} \cdot {\bf j^{\Delta}}\;,
\label{n_Delta}\\ 
& &{\partial {\bf u} \over \partial t} = { {\bf g} \over m n_0} - {{\bf j^{\Delta}}\over n_0}\;,
\label{u} 
\end{eqnarray}
\label{xtalhdyro}
\end{mathletters}
\noindent where $m$ is the mass of the constituent particles and
$n_0$ is the average number 
density. 

These conservation laws must be completed with constitutive equations
for the two fluxes, the momentum flux or  stress tensor, $\sigma_{ij}({\bf r},t)$,
and the defect current, ${\bf j^{\Delta}}({\bf r},t)$.
Assuming purely relaxational dynamics for the defects, 
${\bf j^\Delta}$ is given by
\begin{eqnarray}
{\bf j^{\Delta}}=-\Gamma n_0 {\bf \nabla} p_d \;,
\label{defectcurrent}
\end{eqnarray}
where $\Gamma$ is a constant kinetic coefficient and $p_d$ is the defect pressure, 
\begin{equation}
\label{defectpressure}
p_d=\chi^{-1}\frac{\delta n_{\Delta}}{n_0} 
    + \gamma {\bf \nabla}\cdot{\bf u}\;,
\end{equation}
with $\chi$ a susceptibility  proportional to the sum of the equilibrium 
densities of vacancies and interstitials, and $\gamma$  a 
constant associated with 
the coupling between changes in defect density and bulk 
distortions in the lattice. 
The constitutive relation for the 
stress tensor, $\sigma_{ij}$,  is
\begin{eqnarray}
\sigma_{ij}=2\mu w_{ij}^s + \lambda w_{kk}^s \delta_{ij} + \gamma (\delta n_{\Delta}/n_0) \delta_{ij}  \;,
\label{elasticsigma_ij}
\end{eqnarray}
where $w_{ij}^s=(w_{ij}+w_{ji})/2$ is the symmetric part of the 
strain tensor $w_{ij}$, given by 
\begin{equation}
\label{strain}
w_{ij}=\partial_iu_j \;.
\end{equation}
The coefficients $\mu$ and $\lambda$ are the Lam\'e constants. 
We note that the equilibrium 
bulk modulus, $B_0$, of the solid
is defined in terms of the response to a static isotropic compression
as $B_0=\Big[-\frac{1}{A}\frac{\delta A}{\delta p}\Big]^{-1}$,
with $A$ the area of the crystal.
In the presence of defects it is then given by
\begin{eqnarray}
B_0 = {\mu + \lambda -\gamma^2\chi \over 1 + (\mu+\lambda+2\gamma)\chi}\;,
\label{fullbulkmodulus}
\end{eqnarray}
On the other hand, if one were to measure the bulk modulus on time scales long 
compared to the elastic relaxation time, but short compared to the defect 
diffusion time, the effective bulk modulus, denoted by $B_\infty$, would be 
\begin{eqnarray}
B_{\infty} = \mu + \lambda\;,
\label{elasticbulkmodulus}
\end{eqnarray}
corresponding to the purely elastic contribution obtained in the absence
of defects. In a finite-frequency measurements, the value $B_\infty>B_0$
is reached at sufficiently high frequency. 
Finally, 
changes in the defect density can be related to changes in the overall density and 
bulk strains in the lattice, as 
\begin{eqnarray}
\delta n = \delta n_{\Delta} - n_0 w_{kk} \;,
\label{densitiesrelation}
\end{eqnarray}
where $\delta n=n-n_0$ and $\delta n_{\Delta}=n_\Delta-n^\Delta_0$ 
are the deviations of the
density and defect density  from their equilibrium
values, respectively.

In the presence of dislocations, it is no longer possible to define a continuous
single-valued displacement field, ${\bf u}$. One can define a single-valued,
but discontinuous displacement field by introducing cuts at the 
location of the dislocations. The strain field $w_{ij}=\partial_iu_j$ 
is still single valued and continuous, except at the location of the dislocations,
and it satisfies
\begin{eqnarray}
\epsilon_{ki}\partial_kw_{ij}=a_0 B_j\;,
\label{dislocationcondition}
\end{eqnarray}
where $a_0$ is the lattice spacing,  $\epsilon_{ij}$ the antisymmetric 
unit tensor, and ${\bf B}({\bf r},t)$ is the Burgers-vector charge density.
For a set of $M$ free dislocations at discrete points $\{{\bf R}_\nu\}$,
$\nu=1,2,...,M$, with Burgers vector ${\bf b}_\nu$, it is given by
\begin{equation}
\label{dislocdensity}
{\bf B}({\bf r},t)=\sum_{\nu=1}^M{\bf b}_\nu\delta({\bf r}-{\bf R}_\nu)\;.
\end{equation}
Dislocations are assumed to be annealed and free to equilibrate.
The Burgers vector charge density is an extra hydrodynamic variable with a conservation
law,
\begin{eqnarray}
{\partial B_i \over \partial t} &=& -\partial_j J_j^i\;,
\label{Bconservation}
\end{eqnarray}
where $J_j^i$ is the Burgers current of the $i$th component of dislocation 
charge in the $j$th direction. For length scales longer than the average 
distance between free dislocations, $\xi_d\sim1/\sqrt{n_f}$, and time scales 
longer than the mean collision time between free dislocations, one can 
make some simplifying approximations. Specifically, one can ignore the 
effect of dynamical polarization of bound dislocation pairs and also 
processes by which neutral dislocations pairs 
(i.e., dislocation pairs of opposite Burgers vectors)
are created and annihilated. It was then shown in Ref.~\onlinecite{ZHN},
that the constitutive relation for the Burgers current can be written as
\begin{eqnarray}
J_j^i=-C_{jkl}^i(\sigma_{kl}+p_d\delta_{kl})-D_{jl}^{ik}\partial_lB_k\;.
\label{dislocationcurrent}
\end{eqnarray}
The three terms on the right hand side of Eq.~(\ref{dislocationcurrent})
correspond to the motion of 
dislocations in response to stresses in the lattice (the usual Peach-Koehler 
force), gradients in defect pressure and gradients in the dislocation density, 
respectively. The tensors $C_{jkl}^i$ and $D_{jl}^{ik}$ are given by
\begin{mathletters}
\begin{eqnarray}
C_{jkl}^i &=& \frac{n_f a_0}{4k_B T}\big[(D_g-D_c)\delta_{jk}\epsilon_{li} 
        +(D_g+3D_c)\delta_{ik}\epsilon_{lj}\big]\;,\nonumber\\
\label{C}\\ 
D_{jl}^{ik} &=& \frac{D_g-D_c}{4}(\delta_{ij}\delta_{kl} 
   + \delta_{jk}\delta_{il}) + \frac{D_g+3D_c}{4}\delta_{jl}\delta_{ik}\;,
\nonumber\\
\label{D}
\end{eqnarray}
\label{CDtensors}
\end{mathletters}
where $D_g$ and $D_c$ are the diffusions constants for 
dislocation glide and climb, respectively. 

The conservation laws for density and momentum, Eqs.~(\ref{n}) and (\ref{g}), 
remain unchanged in the presence of dislocations.
A new term must, however, be added to the equation for the conservation 
of defect density,
reflecting the fact that a climbing dislocation acts as a source or sink of 
vacancies and interstitials. Equation (\ref{n_Delta}) is 
then replaced by
\begin{eqnarray}
{\partial n_{\Delta} \over \partial t} &=& -{\bf \nabla} \cdot {\bf j^{\Delta}}+n_0 a_0\epsilon_{ik}J_k^i\;.
\label{adjustedn_Delta}
\end{eqnarray}
Since ${\bf u}$ is multi-valued, it is more convenient to consider
an equation of motion for the strain rather than Eq.~(\ref{u}) 
for the displacement field. This is given by
\begin{eqnarray}
{\partial w_{ij} \over \partial t} &=&\partial_i v_j - {1\over n_0} \partial_i j_j^{\Delta} + a_0\epsilon_{ik}J_k^j\;.
\label{conservationofw_ij}
\end{eqnarray}
where, to linear order, ${\bf g} = mn_0{\bf v}$. 
The first two terms are simply obtained by differentiating
the right hand side of Eq.~(\ref{u}). The third term arises from the 
presence of dislocations. 

The conservation laws of density, Eq.~(\ref{n}), momentum density, 
Eq.~(\ref{g}), defect density, Eq.~(\ref{adjustedn_Delta}), 
and Burgers vector charge, Eq.~(\ref{Bconservation}),
together with the constitutive equations for the currents,
Eqs.~(\ref{defectcurrent}),(\ref{elasticsigma_ij}) 
and (\ref{dislocationcurrent}), and the equation of motion for the local 
strains, Eq.~(\ref{conservationofw_ij}), constitute a closed set of equations
to describe the dynamics of a 2d solid with an equilibrium concentration 
of free annealed dislocations and of vacancy and interstitial defects. 

The work by ZHN focused on the transverse and longitudinal hydrodynamic modes
of these equations. These authors showed that the low frequency modes 
of a solid with dislocations are the
same as those of the hexatic phase. In contrast, our goal here is to 
eliminate the dislocations and defects degrees of freedom and
obtain a closed set of hydrodynamic equations for the density, momentum 
density and bond angle. By this procedure we 
show that the hydrodynamic equations
of a solid with free dislocations are equivalent, at long wavelengths,
to the hydrodynamic equations of a viscoelastic medium with bond angle order.
The latter consists of conservation laws for the density and momentum density,
an equation of motion for the bond angle, with a constitutive equation for the 
stress tensor that takes the form of a differential equation. 

\section{Hydrodynamics of a viscoelastic medium}
\label{grad1}

Our strategy is to eliminate the defect density, $n_{\Delta}$, and the 
Burgers-vector 
charge density, ${\bf B}$, from the hydrodynamic equations of 
a solid with dislocations discussed in the previous section.
The outcome will be 
a closed set of equations for density, momentum 
density, bond angle field and stress tensor. 

It is convenient to write the strain tensor $w_{ij}$
as the sum of its symmetric 
and antisymmetric parts, 
\begin{eqnarray}
w_{ij} &=& (w_{ij}+w_{ji})/2 +(w_{ij}-w_{ji})/2\nonumber\\
       &\equiv& w_{ij}^s + w_{ij}^a\;.
\label{symmantisymmw_ij}
\end{eqnarray}
Only the symmetric part enters in the expression for the stress tensor, 
according to 
Eq.~(\ref{elasticsigma_ij}). One can therefore invert 
Eq.~(\ref{elasticsigma_ij}) to express
$w_{ij}^s$ as a function of the stress tensor and of the density, as
\begin{eqnarray}
w_{ij}^s={1\over2\mu}\sigma_{ij}& & - \delta_{ij}~\frac{\lambda+\gamma}
    {4\mu(\mu+\lambda+\gamma)}~\sigma_{kk} \nonumber\\
    & &- \delta_{ij}~\frac{\gamma}{2(\mu+\lambda+\gamma)}~{\delta n\over n_0}\;,
\label{w_ij^s}
\end{eqnarray}
where Eq.~(\ref{densitiesrelation}) was used to eliminate the defect 
density in favor of the density and the strain.
The antisymmetric part, $w_{ij}^a$, has only one independent
component which is simply the bond angle field, 
$\theta({\bf r},t)$ \cite{footnote},
\begin{eqnarray}
w_{ij}^a=\epsilon_{ij}\theta\;.
\label{theta_dis}
\end{eqnarray}
The equation of motion for the strain, Eq.~(\ref{conservationofw_ij}), 
can then be 
split into an equation of motion for the symmetrized strain, given by
\begin{eqnarray}
\label{conservationofw_ij^s}
{\partial w_{ij}^s \over \partial t} = & & v_{ij} 
    - {1\over 2n_0} \Big(\partial_i j_j^{\Delta}+\partial_j j_i^{\Delta}\Big)
\nonumber\\
    & & + \frac{a_0}{2}\Big(\epsilon_{ik}J_k^j+\epsilon_{jk}J_k^i\Big)\;,
\end{eqnarray}
and an equation for the bond angle field,
\begin{eqnarray}
\frac{\partial\theta}{\partial t}=
        \frac{1}{2}{\bf \hat{z}}\cdot(\nabla\times{\bf v})
   - {1\over 2n_0} \epsilon_{ij}\partial_i j_j^{\Delta}
     + \frac{a_0}{2}J_k^k\;.
\label{bondangle}
\end{eqnarray}
We note that the trace of Eq.~(\ref{conservationofw_ij^s}) simply
gives the continuity equation for the density.

We can now use Eq.~(\ref{w_ij^s}) to rewrite the equation of motion for
the symmetric part of the strain tensor as a differential equation for the time
evolution of the stress tensor,
with the result
\begin{eqnarray}
\frac{\partial\sigma_{ij}}{\partial t}&=&
      2\mu v_{ij}+\lambda\delta_{ij}v_{kk}\nonumber\\
  & &+ a_0\mu\Big(\epsilon_{ik}J_k^j+\epsilon_{jk}J_k^i\Big) +a_0(\lambda+\gamma)\delta_{ij}\epsilon_{kl}J_l^k\nonumber\\
   & &- {\mu\over n_0} \Big(\partial_i j_j^{\Delta}+\partial_j j_i^{\Delta}\Big)
      -\frac{\lambda+\gamma}{n_0}\delta_{ij}\nabla\cdot{\bf j}^\Delta\;.
\label{stress1}
\end{eqnarray}

In order to eliminate the defects and dislocation degrees of freedom, we now
insert the constitutive equations for the defect and dislocation fluxes
on the right hand side of Eqs.~(\ref{bondangle}) and (\ref{stress1})
and use $\delta n_\Delta=\delta n+n_0w^s_{kk}$ and 
$B_j=\epsilon_{ik}\partial_i(w^s_{kj}+w^a_{kj})/a_0$ to eliminate
the defect density and dislocation density
in favor of density and strain field. Finally, the strain field is 
related to the stress tensor and to the bond angle field
using Eqs.~(\ref{w_ij^s}) and (\ref{theta_dis}).
This completes the transformation of the equation for the strain tensor
into two closed coupled equations for the stress tensor 
and the bond angle field.

For clarity, we present in this section the equations obtained when
only
terms of lowest order in the hydrodynamic fields are retained.
In the next  section we will include higher order terms and show
that these yield wavevector dependent
relaxation times for shear and compression. 
At long wavelengths, the gradients of the defect current density
can be neglected both in the equation for the bond angle,
Eq.~(\ref{bondangle}), and in the equation for the stress tensor, 
Eq.~(\ref{stress1}). Similarly, one can drop the contribution to the
dislocation current due to diffusion of dislocations and approximate
\begin{eqnarray}
J_k^j \simeq  -C_{knl}^j(\sigma_{nl}+p_d\delta_{nl})
\end{eqnarray}
or
\begin{eqnarray}
J_k^j \simeq 
  -C_{knl}^j\bigg[& &\sigma_{nl}
    +\delta_{nl}~\frac{\chi^{-1}+\gamma}{2(\mu+\lambda+\gamma)}~\sigma_{kk}
\nonumber\\
  & &  -\delta_{nl}~\frac{\chi^{-1}(\mu+\lambda)-\gamma^2}{\mu+\lambda+\gamma}
       ~\frac{\delta n}{n_0}\bigg]\;.
\end{eqnarray}
Carrying out some tedious algebra, we obtain the desired set 
of hydrodynamic equations, consisting of exact conservation laws 
for density and momentum, the equation for the bond angle field \cite{footnote2},
\begin{eqnarray}
{\partial \theta \over \partial t} =
  {1\over2}{\bf \hat z}\cdot({\bf \nabla}\times{\bf v})
  +\frac{D_g}{2}\nabla^2\theta\;,
\label{thetaEOM}
\end{eqnarray}
and the constitutive equation for the stress tensor,
\begin{eqnarray}
\label{totalstress}
\frac{\partial\sigma_{ij}}{\partial t} & &
  +\frac{1}{\tau_s}\tilde{\sigma}_{ij}
  +\frac{1}{2\tau_b}\delta_{ij}\sigma_{kk}\nonumber\\
 & &   =2\mu\tilde{v}_{ij}+B_\infty\delta_{ij}v_{kk}
    -\frac{B_0}{\tau_b}\delta_{ij}\frac{\delta n}{n_0}\;,
\label{sigma_ij}
\end{eqnarray}
where we recall that $B_\infty=\mu+\lambda$.
The shear and compressional relaxation times (or ``Maxwell'' times)
are given by
\begin{eqnarray}
& & {1\over \tau_s} = n_f\frac{D_g+D_c}{k_BT}\mu a_0^2   \;,
\label{tau_s}\\
& &{1\over \tau_b} = n_f\frac{D_c}{k_BT}\frac{B_{eff}\chi^{-1} a_0^2}{B_0}\;,
\label{tau_b}
\end{eqnarray}
where $B_{eff}=\mu+\lambda-\chi\gamma^2$ is the ``effective'' bulk 
modulus that one would obtain by
integrating out the defect density fluctuations from the free energy. 
\begin{figure}[bth] 
\centering
\setlength{\unitlength}{1mm} 
\begin{picture}(25,160)(0,-32)
\put(-20,-25){\begin{picture}(20,20)(0,0) 
\includegraphics{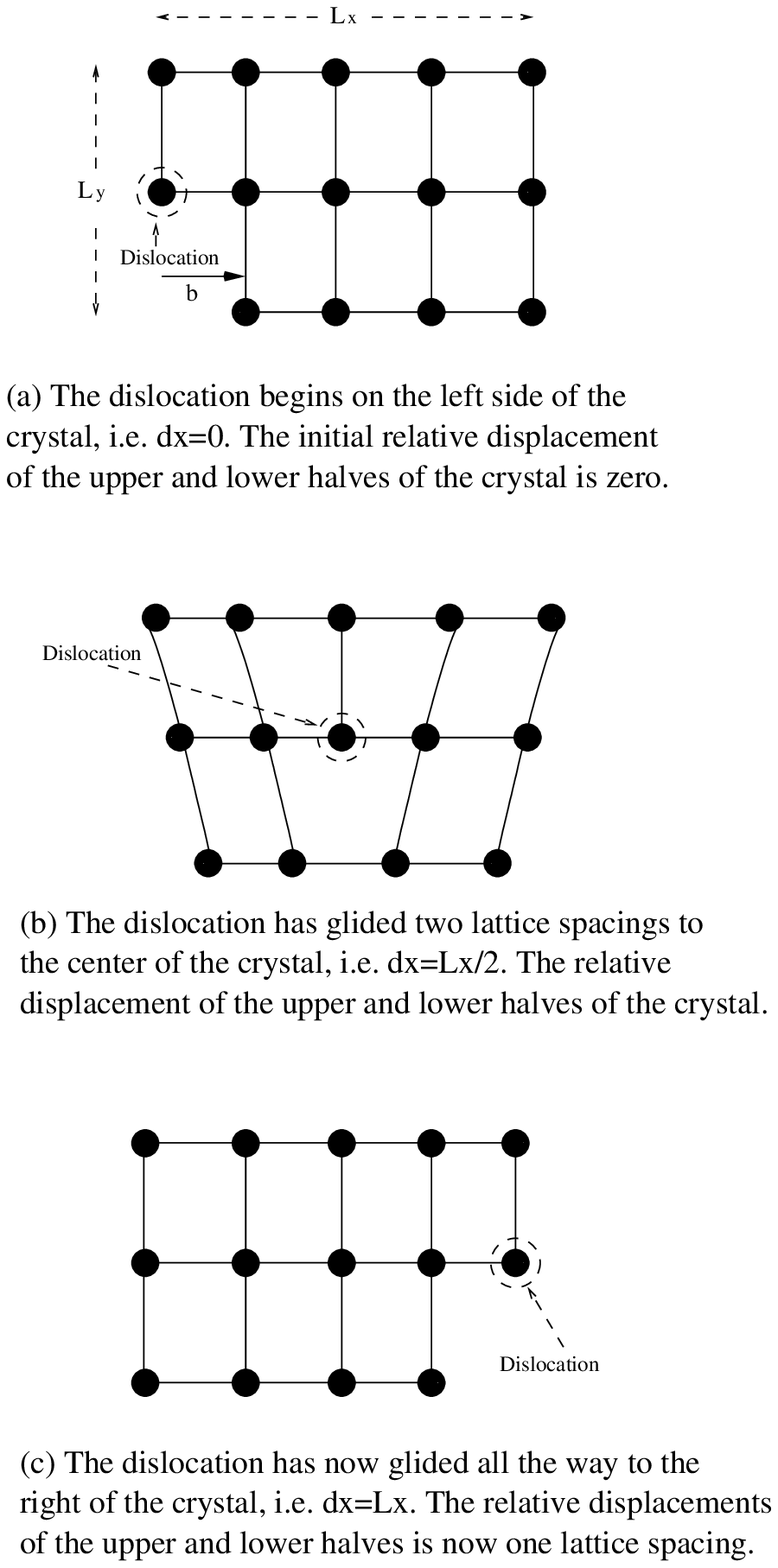} 
\end{picture}} 
\end{picture} 
\caption{An illustration of how dislocation glide leads to relative displacement of two half crystals.}
\label{Orowan} 
\end{figure} 
The shear relaxation time has a simple physical interpretation. 
Considering, for simplicity, only dislocation glide,
the motion of a dislocation of Burgers vector ${\bf b}={\bf \hat{x}}$
a distance $dx$ along its slip line
contributes an amount $\frac{a_0}{L_x}dx$ to the displacement of 
the half crystal,
with $L_x$ the length of the crystal along the slip direction. This is illustrated
for a square lattice in Fig.~(\ref{Orowan}). The corresponding change 
in shear strain is $\sim (a_0dx/L_x)/L_y$, with $L_y$ the width of the crystal
in the direction normal to the slip line. The change in the strain $\gamma$
in a time $dt$
due to $M$ dislocations is then 
$\frac{d\gamma}{dt}\simeq\frac{M}{L_xL_y}a_0\frac{dx}{dt}
=n_fa_0v_g$, with $v_g=\frac{dx}{dt}$ the dislocation glide velocity.
This is the well-known Orowan law \cite{Cottrell} relating the plastic strain
rate to the dislocation velocity. Dislocation motion is driven by local stresses
and we can write $v_g=\mu_g F_s$, with $\mu_g=D_g/(k_BT)$ the 
dislocation mobility and $F_s\approx\sigma_sa_0\approx\mu a_0$ the force on a dislocation
due to the shear stress, $\sigma_s$. If we define a shear relaxation time
by $\frac{\gamma}{dt}\sim\frac{\gamma}{\tau_s}$, and assume $\gamma\sim{\cal O}(1)$,
we immediately obtain $1/\tau_s\sim n_f\frac{D_g}{k_BT}\mu a_0^2$.
When both glide and climb can occur the mobilities are simply additive.

An alternative interpretation of Eq.~(\ref{tau_s}) is as follows.
The rate at which dislocations diffuse
thermally (via glide and climb) a distance of the order
of the typical separation $\sim n_f^{-1/2}$ between free dislocations
is $1/\tau_{th}= (D_g+D_c) n_f$. This is just the mean time for collisions 
between free dislocations. The ratio of the rate of diffusion of dislocations under 
the influence of shear elastic forces to the rate of thermal diffusion 
is of the order of the ratio of a typical shear elastic energy, $\mu a_0^2$ to thermal energy, 
$k_B T$. With this energetic argument we estimate 
$1/\tau_s\approx 1/\tau_{th}(\mu a_0^2/k_B T)$ which is the result given 
in Eq.~(\ref{tau_s}). Interactions
between dislocations are negligible only at time scales longer 
than the mean collision time, $\tau_{th}$ As a result
our model applies only if $\tau_s>\tau_{th}$. 
Since $\mu a_0^2/k_B \sim{\cal O}(T_m)$, the condition 
$\tau_s>\tau_{th}$ simply corresponds 
to being above the melting temperature, $T_m$.

It is not surprising that the bulk memory time is independent of $D_g$, 
since pure compressional strains can only relax
through dislocation climb. Also, since dislocation climb involves a 
change in particle number, $D_c$ is, 
in general, much smaller than $D_g$, so that $\tau_s<<\tau_b$.
In addition, above the melting 
temperature the response to shear distortions is {\em purely} viscoelastic while the response to bulk 
distortions is effectively still elastic at large scales, 
with some dissipative effects (due to the climbing of dislocations)
which, due to the smallness of $D_c$, are expected to be negligible.

In the limit of $n_f\rightarrow 0$, i.e., as $T_m$ is approached from above, 
the shear and compressional relaxation times diverge, 
$\tau_{s,b}\rightarrow\infty$ according to Eq.~(\ref{tau_divergence}), and we obtain
\begin{mathletters}
\begin{eqnarray}
\label{sigma_ijtau->infty}
& &\frac{\partial\tilde{\sigma}_{ij}}{\partial t}
   =2\mu{\tilde v}_{ij}\;,\\
& & {\partial \sigma_{kk}^{\rm dis} \over \partial t}=2(B_{\infty}-B_0)v_{kk}
   =-2{B_{\infty}-B_0\over n_0}{\partial \delta n \over \partial t}\;,
\label{sigma_kk'tau_b->infty}
\end{eqnarray}
\end{mathletters}
\noindent In the absence of dislocations ${v}_{ij}=\partial_t w_{ij}^s$.
Integrating Eq.~(\ref{sigma_ijtau->infty}) then yields the 
shear stress-strain relation for an elastic solid,
\begin{eqnarray}
\tilde{\sigma}_{ij} =2\mu\big(w_{ij}^s-\frac{1}{2}\delta_{ij}w_{kk}^s\big)\;.
\label{sigma_ijelastic}
\end{eqnarray}
Similarly, to lowest order in the gradients, we can neglect  
the rate of change of defect density. Integrating 
Eq.~(\ref{sigma_kk'tau_b->infty}) and using Eq.~(\ref{densitiesrelation}), 
we then obtain $\sigma_{kk}^{\rm dis}=(B_\infty-B_0)w_{kk}^s$ and
\begin{eqnarray}
{\sigma}_{kk} =2B_\infty w_{kk}^s\;.
\label{sigma_kkelastic}
\end{eqnarray}

Far above $T_m$, free dislocations proliferate and both
$\tau_{s}$ and $\tau_b$ become very small. In this limit Eq.(\ref{totalstress}) 
then reduces to 
\begin{eqnarray}
{\sigma}_{ij} =2\mu\tau_s{\tilde v}_{ij}
     +\delta_{ij}(B_\infty-B_0)\tau_b\delta_{ij}\nabla\cdot{\bf v}
     -\delta_{ij}B_0\frac{\delta n}{n_0}\;.
\label{sigma_ijtau->0}
\end{eqnarray}
Identifying the combinations $\mu\tau_s$ with the shear viscosity, $\eta_s$, 
and the combination $(B_\infty-B_0)\tau_b$ with the bulk viscosity, $\eta_b$,
Eq.~(\ref{sigma_ijtau->0})
is just the stress-rate of strain relation of a viscous liquid.

When only terms of lowest order in the gradients of the hydrodynamic fields
are retained, it is not apparent from the viscoelastic stress-rate of strain 
relation that a crystal with annealed dislocations 
is a hexatic, rather than a isotropic viscous liquid.
The presence of quasi-long-range  bond orientational order in the hexatic 
manifests itself in a 
resistance to distortions that produce a non-zero gradient in the bond-angle 
field, $\theta$. Contributions to the hydrodynamic equations from gradients 
of $\theta$ correspond to spatial gradients of $w_{ij}$, which have been 
neglected in this section. Such terms must be retained to see a signature of 
bond orientational order in the stress-rate of strain relations. This is done 
in the next section.

\section{Inclusion of Higher Order Gradients} 
\label{grad2}

In this section we incorporate all diffusive terms 
that were neglected
in the previous section. We do so, however, only for a simplified model
where dislocation climb is neglected. This corresponds to 
freezing out  fluctuations in the defect density. 
It is an excellent approximation in most cases
as the diffusion constant $D_c$ for dislocation climb is generally
very small due to the large energy cost of creating 
vacancies and interstitials. In the absence of fluctuations in the defect density  
the relation between symmetrized strain and stress simplifies to
\begin{eqnarray}
w_{ij}^s={1\over2\mu}\sigma_{ij} - {\lambda\over 4\mu(\mu+\lambda)}\sigma_{kk}\delta_{ij}\;.
\label{simpw_ij^s}
\end{eqnarray}
The exact
conservation laws for density and momentum are of course unchanged and are still given 
by Eqs.~(\ref{n}) and ~(\ref{g}), respectively. The equation of motion for the bond angle 
field now contains terms due to dislocation diffusion, which couple to gradients of stress
and it is given by
\begin{eqnarray}
{\partial \theta \over \partial t} ={1\over2}{\bf \hat z}\cdot({\bf \nabla}\times{\bf v})+
{D_g \over2}\nabla^2\theta-{D_g\over4\mu}\epsilon_{jk}\partial_i\partial_j{\tilde\sigma}_{ij}\;.
\label{thetaEOM3}
\end{eqnarray}
>From Eq.~(\ref{thetaEOM3}), one sees that the presence of gliding dislocations 
leads to a dissipative term $\propto \nabla^2\theta$ and a coupling to 
${\tilde\sigma}_{ij}$. 

In the absence of fluctuations in the defect density,
the stress tensor's trace is simply
\begin{eqnarray}
\sigma_{kk}=-2B_\infty {\delta n \over n_0}\;.
\label{sigma_kk''}
\end{eqnarray}
We therefore focus on its traceless part, $\tilde{\sigma}_{ij}$
Inclusion of the extra diffusive terms in the constitutive relation for ${\tilde\sigma}_{ij}$ 
will  yield terms involving the gradient of the Burgers vector charge density. These terms 
are easily recast as gradients of bond angle and stress, resulting in the following constitutive 
relation for ${\tilde\sigma}_{ij}$
\begin{eqnarray}
{\partial {\tilde\sigma}_{ij} \over \partial t} &+& {{\tilde\sigma}_{ij} \over \tau_s}-{D_g\over4}\nabla^2{\tilde\sigma}_{ij}
\nonumber\\
&=&2\mu{\tilde v}_{ij} + {\mu D_g\over 2}\left(\epsilon_{ik}\partial_j\partial_k+\epsilon_{jk}\partial_i\partial_k\right)\theta \nonumber\\
&+&{\mu D_g\over4}\left(\partial_i\partial_j-{\delta_{ij}\over2}\nabla^2\right){\delta n \over n_0}\;,
\label{sigmatilde_ij'''}
\end{eqnarray}
where  Eq.~(\ref{sigma_kk''}) relating  $\sigma_{kk}$ and $\delta n/n_0$ has been used. It is 
instructive to rewrite Eq.~(\ref{sigmatilde_ij'''}) in Fourier space. The equation for ${\tilde\sigma}_{ij}({\bf k},t)$
is then given by
\begin{eqnarray}
{\partial {\tilde\sigma}_{ij} \over \partial t}  +  {{\tilde\sigma}_{ij} \over \tau_s(k)}
&=& 2\mu{\tilde v}_{ij} -{\mu D_g\over2}\left(\epsilon_{ik}k_jk_k+
\epsilon_{jk}k_ik_k\right)\theta \nonumber\\
&-&{\mu\over\tau_s(0)}\left({\hat k}_i {\hat k}_j-\delta_{ij}/2\right) \xi^2k^2{\delta n\over n_0}\;,
\label{sigmatilde_ijFS1}
\end{eqnarray}
with ${\hat k}_i\equiv k_i/k$ and where we have defined 
\begin{eqnarray}
\xi\equiv{1\over2}\sqrt{D_g\tau_s} \;.
\label{xi1}
\end{eqnarray}
One is naturally led to introduce a  finite wavevector shear relaxation 
time, $\tau_s(k)$,  given by
\begin{eqnarray}
\tau_s(k)= \frac{\tau_s}{1+\xi^2k^2} \;.
\label{tau_s(k)}
\end{eqnarray}
When the constitutive relation is written in this form, it is apparent
that $\xi$ is the length scale beyond 
which the small gradient approximation is valid. 
At low frequency 
Eq.~(\ref{sigmatilde_ijFS1}) reduces to
\begin{eqnarray}
{\tilde\sigma}_{ij} &=& {2\eta_s\over 1+\xi^2k^2}{\tilde v}_{ij} - {\mu D_g\tau_s(0)\over2(1+\xi^2k^2)}\left(\epsilon_{ik}k_jk_k+
\epsilon_{jk}k_ik_k\right)\theta \nonumber\\
&-&{\mu\over 1+\xi^2k^2}\left({\hat k}_i {\hat k}_j-\delta_{ij}/2\right) \xi^2k^2{\delta n\over n_0}\;.
\label{sigmatilde_ijFSdc}
\end{eqnarray}
By comparing Eq.~(\ref{sigmatilde_ijFSdc}) to the corresponding relation for a hexatic given in ZHN,
\begin{eqnarray}
{\tilde\sigma}_{ij}^{hex}=2\eta_s{\tilde v}_{ij} + 
{K_A\over2}\left(\epsilon_{ik}\partial_j\partial_k+\epsilon_{jk}\partial_i\partial_k\right)\theta\;,
\label{hexatic}
\end{eqnarray}
it is natural to identify the combination $\mu D_g\tau_s(0)$ with $K_A$, 
the elastic constant associated with long 
wavelength distortions of the bond angle field.  
The length scale $\xi$ can then be written as
\begin{eqnarray}
\xi\equiv{1\over2}\sqrt{K_A\over \mu} \;
\label{xi2}
\end{eqnarray}
and it represents the length scale below which the energy associated 
with bond angle distortions becomes comparable to that associated with spatially uniform shear 
distortions.

The effective shear relaxation time, 
$\tau_s(k)$, is larger at short length scales, as the 
additional stiffness due to bond orientational order makes the system 
more solid-like.
Finally, it should be noted that the term in  Eq.~(\ref{sigmatilde_ijFS1}) 
$\propto$ $\delta n /n_0$  only contributes to the equation for the
longitudinal ($\propto{\bf \nabla}\cdot{\bf g}$) component of the momentum,
but not to that for the transverse part ($\propto{\bf \nabla}\times{\bf g}$).
This indicates that, even if dislocation climb is forbidden, at short length scales the longitudinal 
component of momentum is affected by
dislocation glide, which yields higher order
pressure-like gradients in the equations of motion. 

Clearly a regime where the wavevector dependence of the shear relaxation time
is important only exists provided $\xi>\xi_d$, with 
$\xi_d\sim1/\sqrt{n_f}$ the average distance between free dislocations.
At length smaller than $\xi_d$ interactions among dislocations can 
no longer be neglected and our model becomes invalid.
It is easy to see that
\begin{eqnarray}
{\xi\over\xi_d}\sim\sqrt{k_BT\over \mu a_0^2} \sim\sqrt{\frac{T}{T_M}}\;,
\label{xiVSxi_d}
\end{eqnarray}
with $T_M$ the melting temperature of the two-dimensional lattice.
Since our work always describes the region $T>T_M$, where free dislocations
are present in the crystal, the wavevector dependence of the relaxation time will 
generally be important.

\vspace{0.2in}
It is a pleasure to acknowledge discussions with Alan Middleton and Jennifer Schwarz. This work was supported by the National Science Foundation
through grant DMR-9805818.

\end{multicols}

\begin{references}
\bibitem{footnote} In a dislocation-free solid, the bond angle field is slaved to the 
displacement field and is given by $\theta = (\partial_x u_y - \partial _y u_x)/2$ 
In the presence of dislocations it can be rewritten as $\theta=\epsilon_{ij}w_{ij}/2$
which, when inverted, gives Eq.~(\ref{theta_dis}).
\bibitem{footnote2} The diffusive term in the equation for 
the time evolution of $\theta$ originates from the dislocation 
diffusion current that was neglected in the other equations. 
This term must, however, be incorporated in the equation for the bond
angle field to obtain
the lowest order term in the relaxation rate of the hexatic mode.
\end{references}
\end{document}